\documentclass[12pt]{article}

\usepackage{amsmath}
\usepackage{amsthm}
\usepackage{amsfonts}
\usepackage{amssymb}
\usepackage{graphicx}
\usepackage{mathrsfs}

\usepackage{url} 

\newcommand{\blind}{0}

\usepackage{algorithmic}

\usepackage{rotating} 

\usepackage{dsfont} 

\usepackage{setspace}

\newcommand{\pd}[2]{\frac{\partial #1}{\partial #2}}
\newcommand{\pdm}[3]{\frac{\partial^{#3} #1}{\partial #2^{#3}}}

\newcommand{\dx}[1]{\ \text{d} #1}
\newcommand{\E}{\mathbb{E}}
\newcommand{\Var}{\text{Var}}

\newcommand{\indicator}[1]{\mathds{1}\{ #1 \}}

\newcommand{\er}{Erd\H{o}s-R\'enyi }

\newcommand{\A}{\mathbf{A}}

\newcommand{\C}{\mathbf{C}}

\newcommand{\w}{\mathbf{w}}

\newcommand{\bu}{\mathbf{u}}
\newcommand{\bd}{\mathbf{d}}
\newcommand{\Y}{\mathbf{Y}}

\newtheorem{defn}{Definition}

\newtheorem{assump}{Assumption}

\usepackage{natbib}
\title{Hidden population size estimation \\ from respondent-driven sampling: a network approach}

\date{}

\begin{document}

\def\spacingset#1{\renewcommand{\baselinestretch}%
{#1}\small\normalsize} \spacingset{1}

\if0\blind
{

\author{Forrest W. Crawford$^1$, Jiacheng Wu$^1$, and Robert Heimer$^2$ \\[0.3em]
\small 1. Department of Biostatistics \\
\small 2. Department of Epidemiology of Microbial Diseases\\
\small Yale School of Public Health }

\maketitle

\noindent \small \textbf{Acknowledgements:} FWC was supported by NIH/NCATS grant KL2 TR000140, NIMH grant P30MH062294, and startup funds from the Yale School of Public Health.  We are grateful to Leonid Chindelevitch, Mark Handcock, Robert Heimer, Edward M. Kaplan, and Li Zeng for helpful comments.  The RDS data presented in the application are from the ``Influences on HIV Prevalence and Service Access among IDUs in Russia and Estonia'' study, funded by NIH/NIDA grant 1R01DA029888 to Robert Heimer and Anneli Uuskula (Co-PIs).  We acknowledge the Yale University Biomedical High Performance Computing Center for computing support, funded by NIH grants RR19895 and RR029676-01.
} \fi

\if1\blind
{
  \author{[Blinded for peer review]}
  \maketitle
} \fi

\bigskip

\begin{abstract}
\noindent Estimating the size of stigmatized, hidden, or hard-to-reach populations is a major problem in epidemiology, demography, and public health research.  Capture-recapture and multiplier methods have become standard tools for inference of hidden population sizes, but they require independent random sampling of target population members, which is rarely possible.  Respondent-driven sampling (RDS) is a survey method for hidden populations that relies on social link tracing.  The RDS recruitment process is designed to spread through the social network connecting members of the target population.  In this paper, we show how to use network data revealed by RDS to estimate hidden population size.  The key insight is that the recruitment chain, timing of recruitments, and network degrees of recruited subjects provide information about the number of individuals belonging to the target population who are not yet in the sample.  We use a computationally efficient Bayesian method to integrate over the missing edges in the subgraph of recruited individuals.  We validate the method using simulated data and apply the technique to estimate the number of people who inject drugs in St.~Petersburg, Russia.
\end{abstract}


\newpage
\spacingset{1.45} 


\section{Introduction}

Estimating the size of stigmatized, hidden, or hard-to-reach populations such as homeless people, sex workers, men who have sex with men, or drug users is an important part of epidemiological, demographic, and public health research \citep{UNAIDS2010Guidelines,bao2010estimating,WHO2014Consolidated,Abdul2014Estimating}.  Census-like enumeration of hidden population members is usually impossible since potential subjects may fear persecution if they participate in a research study.  When random sampling of target population members is feasible, multiplier \citep[e.g.][]{Heimer2010Estimation,Hickman2006Estimating,quaye2015critique} and capture-recapture methods \citep{Fienberg1972Multiple,Laska1988Estimating,Larson1994Indirect,Hall2000How} for estimating population size may perform well.  
Unfortunately random sampling is often impossible because there is no sampling ``frame''; population members are not directly accessible to researchers.  This difficulty has led researchers to develop survey techniques and corresponding statistical tools that do not require random sampling and instead rely on properties of social networks.  

In ``snowball sampling'', subjects enumerate their social contacts, each of whom enters the study, and the process repeats \citep{Goodman1961Snowball}. Since snowball sampling reveals the network (induced subgraph) of respondents, the sample may carry information about global properties of the social network connecting members of the hidden population. \citet{Frank1994Estimating} estimate hidden population size from snowball samples by making homogeneity assumptions about the underlying social network, and \citet{David2002Estimating} use the method to estimate the number of homeless people in Budapest.  Further design-based approaches to population size estimation using snowball sampling have been developed \citep{felix2004combining,Felix2009Link,vincent2012estimating}.  Snowball sampling is often not feasible because social contacts of participants may decline to enroll in the study.  When this happens, the subgraph of respondents may be incomplete, and estimation of population properties -- especially the size of the population -- may suffer.

The network scale-up method is an alternative technique in which researchers survey members of the general population to determine how many people they know (their personal network size), and how many people they know who are members of the target population \citep{Killworth1998Estimation,Bernard2010Counting}.  The proportion of respondents' contacts who are members of the target population is assumed to be equal to the population proportion.  Multiplying this proportion by the known general population size produces an estimate of the target population size.  The network scale-up method has been successfully used to estimate the size of groups at risk of HIV infection, including men who have sex with men, injection drug users, and sex workers \citep{Kadushin2006Scale,salganik2011assessing,Ezoe2012Population,Shokoohi2012Size,Guo2013Estimating}.  The method is appealing because researchers do not need access to the hidden population, but its validity relies on 
subjects' knowledge of their contacts' membership in the target population \citep{Killworth1998Estimation}. 
Sometimes membership in the target population is obscured from non-members \citep{Shelley1995Who,Shelley2006Who}, or groups within the general population may have different probabilities of ties to the target population \citep{Snidero2004Network,Zheng2006How,McCormick2010How,Feehan2014Estimating}.  

Respondent-driven sampling (RDS) is a widely used procedure for recruiting members of hard-to-reach populations for surveys and interventions that relies on participants to recruit other subjects \citep{Heckathorn1997Respondent,Broadhead1998Harnessing}. Beginning with an initial group of participants called ``seeds'', subjects are interviewed and given a reward for participation.  Subjects then receive a small number of ``coupons'' that they can use to recruit other eligible subjects.  Each coupon is marked with a unique ID traceable back to the recruiter.  Subjects recruit others into the study by giving them a coupon that they ``redeem'' by enrolling in the study.  When a new subject enrolls and is interviewed, their recruiter receives a reward.  In this way, the RDS recruitment process is designed to spread through the social network of the hidden population.   One common feature of all RDS surveys is that researchers assess each subject's network degree, the number of other members of the target population the subject knows.  Because of privacy restrictions, subjects typically do not provide identifying information about members of their social network.  Most statistical work on RDS has focused on estimators for population means \citep{Salganik2004Sampling,Volz2008Probability,Gile2011Improved}. 

Does RDS reveal information about the size of the target population?  Just as in snowball sampling and the network scale-up method, subjects report how many members of the target population they know.  Unlike network scale-up surveys, only members of the target population are recruited to participate in an RDS study. In contrast to snowball sampling, not all social contacts of the subject are surveyed: in RDS the subjects decide which of their contacts to recruit.  Despite these limitations, \citet{Paz2011How} use RDS to perform the recapture step of a capture-recapture experiment, even though recruited individuals are not sampled uniformly at random from the target population \citet[see, e.g.][for commentary]{Berchenko2011Capture}.  Recently \citet{Handcock2014Estimating,Handcock2015Estimating} proposed a population size estimator for RDS based on ideas from without-replacement sampling proportional to size \citep{Bickel1992Nonparametric,Gile2011Improved}. Their successive sampling size (SS-size) estimator depends only on the time-ordered sequence of observed network degrees in the RDS sample.  By assuming that RDS is a sampling mechanism that recruits individuals without replacement and with probability proportional to their network degree, \citet{Handcock2014Estimating} and \citet{Handcock2015Estimating} reason that the average degrees of recruited individuals should decrease monotonically with the number of recruited subjects.  The rate of this decrease is believed to reveal information about the size of the population via early depletion of high-degree individuals.  The RDS Analyst software implements the SS-size method \citep{RDSAnalyst}.

In this paper, we take a network-based approach to population size estimation from RDS, based on the intuition behind the snowball sampling estimator and the network scale-up method. The key insight is that the RDS recruitment chain, timing of recruitments, and the degrees of recruited subjects provide information about the number of links between sampled and unsampled population members, and hence the total population size.
We first describe the graphical structure of data obtained from RDS, including the recruitment graph and recruitment-induced subgraph.  The unobserved portions of the recruitment-induced subgraph are treated as missing data.  We describe a Bayesian framework for marginalizing over the missing edges in the recruitment-induced subgraph to estimate population size.  The method relies only on data traditionally obtained by RDS and does not require a change to current RDS recruitment protocol, nor a separate survey of subjects who are not members of the target population. The computational burden of the inference procedure scales with the sample size, not the total hidden population size.  We validate the proposed technique using simulated data and apply the method to estimate the number of injection drug users in St.~Petersburg, Russia.

\section{The graphical structure of RDS data}

\label{sec:graph}

\begin{figure}
  \centering
  \includegraphics[width=0.6\textwidth]{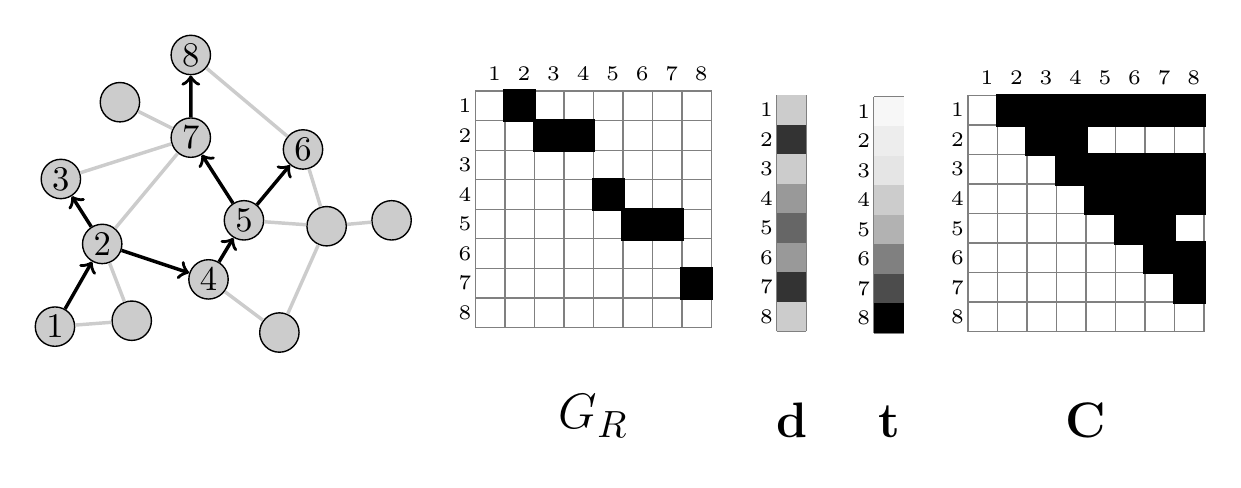}
  \caption{Illustration of the observed data in RDS surveys.  At right, the recruitment graph $G_R$ is shown overlaid on the population graph $G$.  Vertex 1 is a seed, and an arrow from $i$ to $j$ indicates that $i$ recruited $j$.  Several vertices in $G$ remain unsampled.  Researchers conducting an RDS study observe neither $G$ nor the induced subgraph of the sampled vertices.  Next the observed data are shown: the adjacency matrix of the recruitment graph $G_R$, the vector of degrees $\bd$, the vector of recruitment times $\mathbf{t}$, and the coupon matrix $\C$.  The numbered rows and columns correspond to the sampled vertices, numbered in the order of their recruitment.  This figure is adapted from \citet{Crawford2014Graphical}. }
  \label{fig:obs}
\end{figure}

In this section, we outline the observed data in typical RDS surveys of hidden populations, drawn from the definitions given by \citet{Crawford2014Graphical}.  Suppose that the hidden population social network is $G=(V,E)$, where $|V|=N$ is the size of the target population and $G$ contains no self-loops or parallel edges.  A vertex in $G$ is \emph{recruited} if it is known to the study.  A recruited vertex cannot be recruited again. 
\begin{defn}[Recruitment graph]
  The directed recruitment graph is $G_R=(V_R,E_R)$, where $V_R\subset V$ is the set of $n$ sampled vertices and a directed edge $\{i,j\}\in E_R$ indicates that $i$ recruited $j$.
\end{defn}
\noindent Since subjects cannot be recruited more than once, $G_R$ is acyclic.  
\begin{defn}[Degree]
  A vertex's degree is the number of edges incident to it that connect to vertices in the hidden population graph $G$.
\end{defn}
\begin{defn}[Recruitment-induced subgraph] 
 The recruitment-induced subgraph is an undirected graph $G_S=(V_S, E_S)$, where $V_S=V_R$ consists of $n$ sampled vertices, and $\{i,j\}\in E_S$ if and only if $i\in V_S$, $j\in V_S$, and $\{i,j\}\in E$.
\end{defn}
\noindent Let $\bd$ be the time-ordered $n\times 1$ vector of subjects' degrees in the order they were recruited into the study and let $\mathbf{t} = (t_1,\ldots,t_n)$ be the $n\times 1$ vector of recruitment times, where $t_1<\cdots<t_n$.  
\begin{defn}[Coupon matrix]
Let $\C$ be the $n\times n$ coupon matrix whose element $\C_{ij}$ is 1 if subject $i$ has at least one coupon just before the $j$th recruitment event, and zero otherwise.  The rows and columns of $\C$ are ordered by subjects' recruitment time.  
\end{defn}
\noindent The observed data from the RDS recruitment process is $\Y=(G_R,\bd,\mathbf{t},\C$).
Figure \ref{fig:obs} illustrates the observed data and their relationship to the unobserved population graph $G$.  Since the recruitment graph $G_R$ does not contain any edges along which a recruitment event did not take place, the recruitment-induced subgraph $G_S$ is not fully observed. However, the observed degrees $\bd$ and the edges in the recruitment graph $G_R$ place restrictions on the number of non-recruitment edges that can connect vertices in $V_S$, and it is intuitively clear that an estimate $\widehat{G}_S$ of $G_S$ must adhere to certain compatibility conditions.
\begin{defn}[Compatibility] 
  \label{def:compatibility}
  An estimated subgraph $\widehat{G}_S=(\widehat{V}_S,\widehat{E}_S)$ is compatible with the observed data $(G_R,\bd)$ if the following conditions are met: 1. the vertices in the estimated subgraph are identical to the set of recruited vertices: $v\in\widehat{V}_S$ if and only if $v\in V_R$; 2. all directed recruitment edges are represented as undirected edges: for all $(i,j)\in E_R$, $\{i,j\}\in\widehat{E}_S$; 3. the number of edges in $G_S$ belonging to each sampled vertex does not exceed the vertex's degree: for all $v\in V_R$, $\sum_{u\in V_R\setminus v} \indicator{\{u,v\}\in \widehat{E}_S} \leq d_v$, where $d_v$ is the degree of vertex $v$.  
\end{defn}
\noindent These compatibility conditions provide topological constraints on the structure of $\widehat{G}_S$.  


\section{Inference for the population size}

\label{sec:inference}

\begin{figure}
  \centering
  \includegraphics[width=0.6\textwidth]{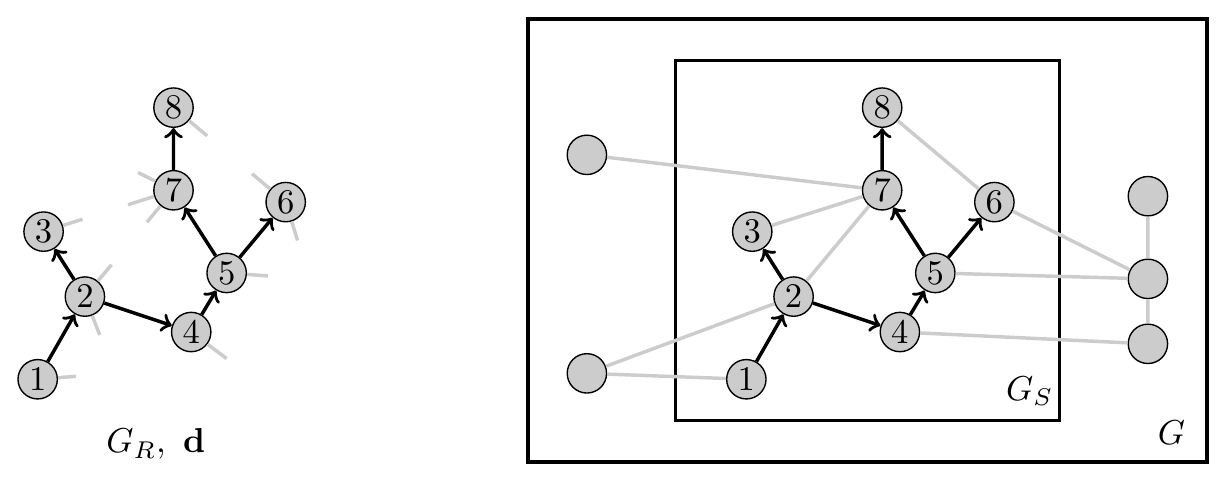}
  \caption{Illustration of population size estimation task using the graphical structure of data obtained by RDS.  We seek the number of vertices $N$ in the graph $G=(V,E)$.  The observed recruitment graph $G_R$ is shown at left.  Each vertex is augmented with the number of pendant edges implied by its degree.  RDS data do not directly reveal which of these pendant edges connect to observed vertices, and which connect to unobserved vertices.  At right, the recruitment-induced subgraph $G_S$ has been reconstructed, revealing the number of edges that connect to unsampled vertices at each step of the recruitment process.  Section \ref{sec:er} provides a derivation of the likelihood of $N$ given $G_S$. }
  \label{fig:popsize}
\end{figure}

In this section, we construct a probability model by which the observed data $\Y=(G_R,\bd,\mathbf{t},\C)$ in an RDS survey are linked to the number of vertices $N$ in the target population.  Figure \ref{fig:popsize} illustrates the problem of estimating the number of vertices $N$ in $G$ from the recruitment-induced subgraph $G_S$.  First we show that if the recruitment-induced subgraph $G_S$ is known, a simple statistic -- the number of pendant edges connecting each sampled vertex to unsampled vertices at the moment of recruitment -- can be used to derive the likelihood of $N$ conditional on $G_S$.  Next, we appeal to results by \citet{Crawford2014Graphical} giving the likelihood of the recruitment-induced subgraph $G_S$ and a per-edge recruitment rate parameter $\lambda$.  Our strategy is to marginalize over $G_S$ and $\lambda$ to arrive at the posterior distribution of $N$.

\subsection{Likelihood of $N$ given $G_S$ under the \er model}

\label{sec:er}

We first state some assumptions about the social network connecting members of the hidden population and the RDS recruitment process on this network.  
\begin{assump}[Existence of a network]
  The target population social network is a finite graph $G=(V,E)$ with no parallel edges or self-loops.
  \label{assump:network}
\end{assump}
\noindent Network-based methods for population inference must make homogeneity assumptions to ensure that a sub-sample of the network can be used to make inference about the total network.  In the \er random graph model, each edge between vertices is formed independently with probability $p$ \citep{Erdos1959Random,Erdos1960Evolution}.  Let $G\sim\mathcal{G}(N,p)$ denote an \er random graph.  The degree $d_i$ of a vertex $i$ has distribution 
\begin{equation}
  d_i\sim \text{Binomial}(N-1,p) ,
  \label{eq:di}
\end{equation}
where $N=|V|$.  The likelihood of a particular graph $G$ depends only on the number of edges $|E|$,  
\begin{equation}
L(N,p|G) = p^{|E|} (1-p)^{\binom{N}{2} - |E|} .
\label{eq:erlik1}
\end{equation}
The \er random graph model formalizes the notion of independent and identically distributed (with probability $p$) formation of reciprocal social ties between individuals in a finite population.  While the \er model is believed to be a poor generative model for non-hidden social networks \citep{Watts1998Collective,Robins2001Network}, very little is known about the structure of contacts between members of highly stigmatized or criminalized populations.  The \er model has proven to be empirically very useful for estimating hidden population sizes: both the snowball sampling estimator \citep{Frank1994Estimating} and the network scale-up estimator \citet{Killworth1998Estimation} rely on equivalent network homogeneity assumptions.  The real-world usefulness of the \er model for hidden population size estimation suggests that an approximate relationship between individual degrees and the population size $N$ like \eqref{eq:di} may hold in some real-world populations.  

We now specify the distribution of the hidden population graph.  
\begin{assump}[Network model] \label{assump:er}
The target population graph has \er distribution, $G\sim\mathcal{G}(N,p)$.  
\end{assump}
\noindent The likelihood of $G_S$ conditional on $G_R$ and $\bd$ under the \er model depends on assumptions about the dynamics of the RDS recruitment process.  But there is significant disagreement about how to model the recruitment process \citep{Salganik2004Sampling,Gile2010Respondent,Gile2011Improved,Berchenko2013Modeling,Crawford2014Graphical,Malmros2014Respondent}. We therefore make a simple assumption that permits calculation of the distribution of a statistic of $G_S$ under Assumption \ref{assump:er}.  Call a vertex a \emph{recruiter} if it has at least one coupon and shares an edge with an unrecruited vertex.  Call a vertex \emph{susceptible} to recruitment if it has not yet been recruited and shares an edge with a recruiter.  
\begin{assump}[RDS sampling probabilities] \label{assump:samplingprob}
The next recruited vertex is chosen from among all susceptible vertices with probability that depends only on the edges it shares with recruiters.  The edges connecting the newly recruited vertex to other unrecruited vertices do not affect its probability of being recruited.  
\end{assump}
\noindent Assumption \ref{assump:samplingprob} provides a connection between the recruitment probability for each vertex and the structure of the network.  

Under Assumptions \ref{assump:er} and \ref{assump:samplingprob}, the recruitment-induced subgraph $G_S$ is not an \er graph because new vertices may not be chosen uniformly at random from the set of unrecruited vertices.  However, since recruitment probability does not depend on edges not connected to active recruiters, it does not depend on edges connecting unrecruited vertices to other unrecruited vertices in particular.  This intuition yields a suitable probability model linking the subgraph $G_S$ to the population size $N$.  Suppose $G_S$ is known and let $d_i^u$ be the number of edges belonging to vertex $i$ that connect to unknown vertices \emph{at the moment $i$ is recruited} (recall that the indices $i$ are ordered by the time of entry into the study),
\begin{equation}
  d_i^u = \bd_i - \sum_{j=1}^{i-1} \indicator{ \{i,j\}\in E_S} . 
  \label{eq:diudef}
\end{equation}
Then by independence of edges in the \er model, 
\begin{equation}
 d_i^u \sim \text{Binomial}(N-i,\ p) 
 \label{eq:diu}
\end{equation}
unconditional on $d_i$ and $d_j$ for $j\neq i$ and independently of $d_j^u$ for $j\neq i$.  In words, the number of edges connecting a recruited vertex to unrecruited vertices (at the moment it is recruited, before observing its total degree) depends only on the number of remaining unrecruited vertices and $p$.  These $d_i^u$ connections are formed independently with probability $p$ and without replacement to any of the $N-i$ remaining unsampled vertices.

The presence of the population size parameter $N$ in \eqref{eq:diu} suggests that the sequence of $d_i^u$'s may contain information about $N$.  Since $d_1^u,\ldots,d_n^u$ are independent binomial random variables, the joint likelihood of $N$ and $p$, given $\widehat{G}_S$ and $\bd$, is 
\begin{equation}
  L(N,p|\widehat{G}_S,\Y) = \prod_{i=1}^n \binom{N-i}{d_i^u} p^{d_i^u}(1-p)^{N-i-d_i^u},
 \label{eq:LNp}
\end{equation}
where $d_i^u$ is calculated from $\widehat{G}_S$ and $\bd$ via \eqref{eq:diudef}. 



\subsection{Likelihood of $G_S$}

The likelihood \eqref{eq:LNp} permits estimation of $N$, conditional on observation of the recruitment-induced subgraph $G_S$. However, $G_S$ is not directly revealed by the observed data $\Y$. The statistic $\mathbf{d}^u=(d_1^u,\ldots,d_n^u)$ is sufficient for $N$ and $p$, but the graphical structure of $G_S$ induces complex combinatorial dependencies in the elements of $\mathbf{d}^u$, and the marginal probability distribution of $\bd^u$ cannot be represented in a simple way.  We therefore seek a probability model for $G_S$ given $\Y$, and marginalize over the unobserved portion of this graph with respect to this model.  The compatibility conditions given in Definition \ref{def:compatibility} place strong restrictions on the structure and density of $G_S$.  Let $\mathcal{C}(G_R,\bd)$ denote the set of all recruitment-induced subgraphs that are compatible with the observed data $G_R$ and $\bd$.  

The least restrictive option is to marginalize over $G_S$ with respect to the uniform distribution on $\mathcal{C}(G_R,\bd)$ by setting $\pi(G_S)\propto 1$ for $G_S\in \mathcal{C}(G_R,\bd)$ and zero otherwise.  However, the uniform distribution over $\mathcal{C}(G_R,\bd)$ does not give rise to the uniform distribution over $|E_S|$, and most subgraphs in $\mathcal{C}(G_R,\bd)$ have far more edges than the true subgraph $G_S$. The result is that the uniform distribution over subgraphs $G_S$ results in a highly informative distribution over $\bd^u$ that does not place most of its mass near the true value of $\bd^u$.  Alternatively, we could place a prior distribution over the number of edges $|E_S|$ in $G_S$.  To illustrate, let 
  $\pi(G_S)\propto \exp[-\gamma |E_S|]$
for $G_S\in\mathcal{C}(G_R,\bd)$ and zero otherwise. Choosing $\gamma>0$ penalizes dense subgraphs, and all subgraphs with a given number of edges have the same probability under this model.  

A more sophisticated marginalizing distribution can be derived from the time series of recruitment events.  By making assumptions about the time dynamics of the recruitment process on $G_S$, we can calculate the likelihood of the observed recruitment times $\mathbf{t}$ conditional on $G_S$ to develop a probability model for $G_S$.  The recruitment model depends on the following assumptions, drawn directly from \citet{Crawford2014Graphical}. 
\begin{assump}
  Vertices become recruiters immediately upon entering the study and receiving one or more coupons. They remain recruiters until their coupons or susceptible neighbors are depleted, whichever happens first.
  \label{assump:recruiter}
\end{assump}
\noindent Call an edge in $G$ \emph{susceptible} if it links a recruiter and a susceptible vertex.  
\begin{assump}
When a susceptible neighbor $j$ of a recruiter $i$ is recruited by any recruiter, the edge connecting $i$ and $j$ is immediately no longer susceptible.
\end{assump}
\begin{assump}[Exponential waiting times]
The time to recruitment along an edge connecting a recruiter to a susceptible neighbor has exponential distribution with rate $\lambda$, independent of the identity of the recruiter, neighbor, and all other waiting times.  
  \label{assump:exp}
\end{assump}
\noindent Assumptions \ref{assump:recruiter}-\ref{assump:exp} are consistent with Assumption \ref{assump:samplingprob} \citep[for proof, see Propositions 1 and 2 of][]{Crawford2014Graphical}.  

The likelihood of the recruitment time series on a fixed graph can be computed under this model.  Let $\w=(0,t_1-0,t_2-t_1,\ldots,t_n-t_{n-1})$ be the vector of inter-recruitment waiting times.  Let $\A$ be the adjacency matrix of $G_S$, where the rows and columns of $\A$ correspond to vertices in the order of their recruitment into the study.  Let $\mathbf{u}$ be the $n\times 1$ vector whose $i$th element is the number of pendant edges emanating from $i$ to unsampled vertices, $\bu_i=\bd_i-\sum_{j=1}^n \A_{ij}$.  Then the joint likelihood of $G_S$ and the waiting time parameter $\lambda$ is given by
\begin{equation}
  L(G_S,\lambda|\Y) = \left(\prod_{j\notin M} \lambda \mathbf{s}_j\right) \exp[-\lambda \mathbf{s}'\w] ,
  \label{eq:tslik}
\end{equation}
where 
\begin{equation}
  \mathbf{s}=\text{lowerTri}(\A\C)'\mathbf{1} + \C'\bu
  \label{eq:s}
\end{equation}
and $M$ is the set of seeds \citep{Crawford2014Graphical}. Information from the subgraph $G_S$ enters the likelihood through the vector $\mathbf{s}$, the number of susceptible vertices just before each recruitment event.  




\subsection{Posterior distribution of $N$}

We now combine the likelihood expressions \eqref{eq:LNp} and \eqref{eq:tslik} with  prior information to formulate the posterior distribution of $N$.  The joint likelihood is $L(N,p,G_S,\lambda|\Y) =\allowbreak L(N,p|G_S,\Y)\times \allowbreak L(G_S,\lambda|\Y)$.  Assume $N$, $p$, $G_S$, and $\lambda$ are \emph{a priori} independent with prior distributions $\pi(N)$, $\pi(p)$, $\pi(G_S)$, and $\pi(\lambda)$ respectively.  The posterior distribution of $N$ is obtained by marginalizing over compatible subgraphs, $p$, and $\lambda$,
\begin{equation}
  \Pr(N|\Y) \propto \pi(N) \sum_{\widehat{G}_S} \pi(\widehat{G}_S) \int_0^\infty  L(\widehat{G}_S,\lambda|\Y)\ \pi(\lambda) \int_0^1 L(N,p|\widehat{G}_S,\Y)\ \pi(p) \dx{p} \dx{\lambda} .
\label{eq:post1}
\end{equation}
where the sum is over compatible subgraphs $\widehat{G}_S\in\mathcal{C}(G_R,\bd)$.
Let $p$ have Beta$(\alpha,\beta)$ distribution with density $\pi(p) \propto p^{\alpha-1}(1-p)^{\beta-1}$. Let $\lambda$ have Gamma$(\eta,\xi)$ distribution with density $\pi(\lambda)\propto\lambda^{\eta-1} e^{-\xi\lambda}$.  Then integrating analytically over $p$ and $\lambda$ in \eqref{eq:post1}, the posterior distribution of $N$ becomes 
\begin{equation}
\Pr(N|\Y) \propto \pi(N) \sum_{\widehat{G}_S} \frac{\pi(\widehat{G}_S)\prod_{j\notin M} \mathbf{s}_j}{(\mathbf{s}'\w + \xi)^{n-m+\eta}} \left[ \prod_{i=1}^n \textstyle{\binom{N-i}{d_i^u}} \right] \textstyle{\mathrm{B}\left(D^u+\alpha,nN-\binom{n+1}{2}-D^u+\beta\right)}
  \label{eq:post2}
\end{equation}
where $\mathrm{B}(\cdot,\cdot)$ is the Beta function and $D^u=\sum_{i=1}^n d_i^u$ and $\mathbf{s}$ are computed using $\widehat{G}_S$.  A derivation of \eqref{eq:post2} is given the Supplementary Materials.

\subsubsection{Prior distributions for $N$}

Not every value of $N$ is feasible: since no parallel edges are allowed under Assumption \ref{assump:network}, $N$ must be large enough to accommodate all the edges emanating from sampled vertices.  Therefore, we need $N\geq n + \max_i d_i^u$ for the $d_i^u$'s derived from a particular subgraph $\widehat{G}_S$.  Rather than make the prior $\pi(N)$ conditional on each particular realization of $\widehat{G}_S$, we note that $d_i^u\leq d_i$ for every compatible $\widehat{G}_S$ and impose the simpler constraint $N \geq n + \max_{i} d_i$, which does not depend on any particular $\widehat{G}_S$. For surveys where $N\gg n$, this should not pose a problem for estimation of $N$.  Setting $N_\text{min}=n + \max_i d_i$, we will always consider \eqref{eq:post1} to be defined only for $N\ge N_\text{min}$.

A relatively uninformative class of prior distributions for $N$ is the power-law mass function $\pi(N) \propto N^{-c}$ where $c\geq 0$ and $N\geq N_\text{min}$.  When $c>1$ the prior density is proper: $\sum_{N=N_\text{min}}^\infty \pi(N) < \infty$.  When $c>2$ the prior mean exists, and when $c>3$ the prior variance exists.  However, researchers may prefer not to specify a strongly informative prior for $N$, and $c=1$ is a popular choice \citep{Draper1971Bayesian,Raftery1988Inference}.  Unfortunately the posterior distribution $\Pr(N|\Y)$ may not behave well for some values of $c$: \citet{Kahn1987Cautionary} warns that estimates based on the beta-binomial distribution can have undesirable properties under some priors $\pi(N)$.  In the Supplementary Materials, we show that the posterior mass function \eqref{eq:post1} is a proper probability distribution when $\alpha+c>1$; when $\alpha+c>2$ the posterior mean is finite, and when $\alpha+c>3$ the posterior variance is finite.  When posterior moments of interest do not exist, it may be tempting to posit $N_\text{max}$, the largest permissible estimate of $N$, and letting $\pi(N) \propto N^{-c} \indicator{N_\text{min}<N<N_\text{max}}$.  But since the posterior moments for unbounded $N$ are undefined, their estimates under the truncated prior depend acutely on the choice of $N_\text{max}$ and are less influenced by the observed data \citep{Kahn1987Cautionary}.  We therefore consider below specifications of $\pi(N)$ such that the prior has infinite support, the posterior is proper, and at least the first two moments exist.  While this inevitably results in a more informative set of priors, it seems a small price to pay for finite posterior mean and variance.



\subsection{Monte Carlo sampling}

The posterior distribution \eqref{eq:post1} is obtained by marginalizing over compatible subgraphs $\widehat{G}_S$. Under the compatibility conditions in Definition \ref{def:compatibility}, this sum cannot be performed analytically and the distribution of $N$ conditional on $\widehat{G}_S$ does not have a standard form.  We therefore resort to Gibbs sampling: first we sample $\widehat{G}_S$ conditional on $N$, then sample $N$ conditional on $\widehat{G}_S$.  Sampling $\widehat{G}_S$ is remarkably efficient because update expressions are available for the statistic $\mathbf{s}$ in the likelihood \eqref{eq:tslik}, making the matrix multiplications in \eqref{eq:s} unnecessary. Integration over compatible subgraphs $\widehat{G}_S$ is accomplished by proposing changes to the connectivity of $G_S$, then using a Metropolis-Hastings step to accept or reject the proposal.  Sampling $N$ given $\widehat{G}_S$ relies on a close approximation to the conditional distribution. The Supplementary Materials provide a comprehensive description of the Gibbs sampling routine.


\section{Validation using simulated data}

We performed simulations to validate the proposed method for population size estimation from RDS data under the model outlined in Section \ref{sec:inference}.  First, we simulate an \er population network $G=(V,E)$ with $|V|=N=1000$, 5000, and 10000, $p=5/N$, $10/N$, and $15/N$.  Conditional on the simulated population graph, we simulate the RDS recruitment process under typical real-world study conditions with $n=500$ recruitments starting from $|M|=10$ seeds, and three coupons per recruit using the model described by \citet{Crawford2014Graphical}.  From the simulated recruitment data, we extract $\Y=(G_R,\bd,\mathbf{t},\C)$ and estimate the posterior distribution of $N$ given $\Y$ outlined above.  

We employ weakly informative priors for the unknown parameters.  We assign to $N$ the vague improper prior distribution $\pi(N)\propto N^{-1}$.  For the edge density $p$ we assign $p \sim \text{Beta}(\alpha,\beta)$, with $\alpha>2$ and $\beta = \alpha(1-p_\text{true})/p_\text{true}$, where $p_\text{true}$ is the true value of $p$.  This specification ensures that the posterior distribution of $N$ has finite second moment and the prior expectation of $p$ is equal to $p_\text{true}$. 
To evaluate the sensitivity of posterior estimates to variation in the prior parameters, we set $\alpha=3$, 10, and 20; we set the prior variance for $\lambda$ to $v_\lambda=1$, since simulation results appear to be insensitive to the prior variance for $\lambda$.  The prior for $G_S$ is $\pi(G_S)\propto \exp[-\gamma|E_S|]$ where $\gamma=-\log\big(p_\text{true}/(1-p_\text{true})\big)$. For the waiting time parameter $\lambda$, we specify $\eta$ and $\xi$ to give prior mean equal to the true value $\lambda_\text{true}$ and prior variance $v_\lambda$.  Then we let $\eta=\lambda_\text{true}^2/v_\lambda$ and $\xi=\lambda_\text{true}/v_\lambda$, which gives $\E[\lambda] = \lambda_\text{true}$ and $\Var[\lambda]=v_\lambda$.  The true value is $\lambda_\text{true}=1$ for all simulations.  

For each parameter combination, we simulated 100 independent networks and RDS datasets, and for each dataset, we estimate $N$ via its posterior mean.  Table \ref{tab:sim} shows posterior summaries for the simulated data.  For each set of 100 simulations, the true $N$, the expected degree $Np$, and $\alpha$ are shown.  We report the mean of all 100 posterior means, the standard deviation (SD) of the posterior means, and the relative bias $(\E[N|\Y]-N_\text{true})/N_\text{true}$.  Posterior means of the population size $N$ show low, mostly positive bias. The relative bias does not seem to increases slowly with $N$.  The standard deviation (SD) of posterior means increases with higher $N$.  Estimates of $N$ exhibit least bias when $\alpha$ is large, indicating greater certainty about the edge density $p$.  With $\alpha=3$, the posterior mean exists, but the posterior variance does not. This setting explores estimation under the weakest possible prior assumptions about $p$ that nevertheless guarantee that the posterior mean exists.  While it seems from Table \ref{tab:sim} that some bias is present when $\alpha=3$, it is encouraging that relatively weak prior assumptions can still give rise to reasonable estimates.

\begin{table}
  \centering
  \begin{tabular}{rrrrrrrr}
    \hline
    \multicolumn{2}{c}{Simulation}  &&       && \multicolumn{3}{c}{Posterior } \\ 
    \multicolumn{2}{c}{Parameters}  && Prior && \multicolumn{3}{c}{Estimates of $N$} \\ \cline{1-2} \cline{4-4} \cline{6-8}
     $N$ & $Np$  && $a$  && Mean & SD & rel bias \\ 
  \hline
  1000 & 5  && 3  && 1010 & 104 & 0.010 \\ 
       &    && 10 && 1011 & 97 & 0.011 \\ 
       &    && 20 && 1010 & 90 & 0.010 \\ 
       & 10 && 3  && 964  & 67 & -0.036 \\ 
       &    && 10 && 953  & 62 & -0.047 \\ 
       &    && 20 && 967  & 63 & -0.033 \\ 
       & 15 && 3  && 952  & 54 & -0.048 \\ 
       &    && 10 && 950  & 52 & -0.050 \\ 
       &    && 20 && 953  & 51 & -0.047 \\ 
  5000 & 5  && 3  && 5953 & 3208 & 0.191 \\ 
       &    && 10 && 5416 & 1664 & 0.083 \\ 
       &    && 20 && 5126 & 1091 & 0.025 \\ 
       & 10 && 3  && 7072 & 4071 & 0.414 \\ 
       &    && 10 && 5327 & 1495 & 0.065 \\ 
       &    && 20 && 5093 & 1038 & 0.019 \\ 
       & 15 && 3  && 6134 & 2742 & 0.227 \\ 
       &    && 10 && 5434 & 1421 & 0.087 \\ 
       &    && 20 && 5109 & 999 & 0.022 \\ 
 10000 & 5  && 3  && 14240 & 8188 & 0.424 \\ 
       &    && 10 && 10829 & 3536 & 0.083 \\ 
       &    && 20 && 10404 & 2356 & 0.040 \\ 
       & 10 && 3  && 14931 & 9114 & 0.493 \\ 
       &    && 10 && 10932 & 3426 & 0.093 \\ 
       &    && 20 && 10519 & 2357 & 0.052 \\ 
       & 15 && 3  && 13105 & 7010 & 0.311 \\ 
       &    && 10 && 10767 & 3372 & 0.077 \\ 
       &    && 20 && 10415 & 2302 & 0.042 \\ 
   \hline
  \end{tabular}
  \caption{Simulation results with RDS sample size $n=500$, $|M|=10$ seeds, $\lambda=1$, and three coupons per subject. The true value $N$, the expected degree $Np$, and prior parameter $\alpha$ are shown for each set of 100 simulations.  We report the average posterior mean, average posterior SD, and relative bias.  Posterior means of $N$ exhibit low (but mostly positive) bias, with higher SD, as the true population size increases.  Estimates are most accurate when prior information about $p$ is strong, and $\alpha$ is large. }
  \label{tab:sim}
\end{table}






%



\section{Application: how many people inject drugs in St.~Petersburg?}

\label{sec:application}

The Russian Federation has experienced simultaneous epidemics of drug abuse and HIV infection since the mid-1990s, and HIV prevalence is highest among people who inject drugs (PWID) \citep{Abdala2003Estimating,rhodes2004hiv,WHO2005Russian,Pokrovsky2010Hiv,Unaids2010Global}.  Drug possession in Russia can result in serious legal penalties, including incarceration, loss of employment, and revocation of driving privileges.  HIV-positive people in Russia are often subject to strong social stigma and may lack access to treatment and education resources \citep{Balabanova2006Stigma,Sarang2012Systemic,burke2015tale}.  In St.~Petersburg, Russia, HIV incidence and prevalence are high among PWID \citep{Kozlov2006HIV,Niccolai2011Estimates}, and researchers have found that many PWID do not have ready access to HIV testing and are not aware of their HIV status \citep{Niccolai2010High}.  PWID in Russia often obtain drugs through local social networks connecting drug dealers and buyers \citep{Shaboltas2006HIV,Cepeda2011Drug}. The social nature of the drug scene in St.~Petersburg creates problems for public health and epidemiological research on PWID (also called injection drug users -- IDUs):  ``Such a structure makes it difficult to recruit through outreach and easier to recruit by allowing IDUs to penetrate their own network of contacts'' \citep[][page 662]{Shaboltas2006HIV}.  PWID in St.~Petersburg therefore constitute an epidemiologically important hidden population, connected by a social network, for which random sampling is impossible.  

Knowledge of the size of the PWID population in St.~Petersburg would substantially illuminate the number of people at risk for HIV infection, and could help determine the scale and scope of education, treatment, and intervention programs in that community.  To estimate the number $N$ of PWID in St.~Petersburg, \citet{Heimer2010Estimation} use a multiplier method with estimated HIV prevalence (from a different RDS study), HIV testing frequency, and other sources of information to obtain $\hat{N}=83118\pm 5799$.  Given that nearly all epidemiological research on PWID in St.~Petersburg uses RDS to recruit participants, a method for estimating population size directly from RDS data would be particularly useful.  

We analyze data from an RDS study of PWID in St.~Petersburg performed during 2012-2013. Researchers recruited $n=813$ PWID using 17 seeds and conducted interviews to gauge perceived barriers to use of HIV prevention and treatment services.  While the study was not intended to be used for population size estimation, its size and adherence to the traditional RDS recruitment protocol outlined by \citet{Heckathorn1997Respondent} make it an appealing opportunity for population size estimation.  \citet{Crawford2014Graphical} shows the observed data $\Y=(G_R,\mathbf{d},\mathbf{t},\mathbf{C})$ from this study and describes the recruitment procedure in detail.  

We investigate estimation of $N$ under the vague prior $\pi(N)\propto N^{-1}$, $\lambda\sim\text{Gamma}(\eta=1,\xi=1)$, and several specifications for $\pi(p)$, indexed by the parameters $\alpha$ and $\beta$.  To find a suitable prior for $p$ that takes into account both the previous population size estimate of \citet{Heimer2010Estimation} and the requirement that the first two moments of the posterior distribution exist, we adopt an empirical Bayes approach.  In the Supplementary Materials, we describe a method for prior elicitation using a lower bound for $p$, given a prior estimate $\hat{N}$ of $N$.  For the St.~Petersburg data, we find that this bound is $\hat{p}_\text{lo}=1.26\times 10^{-5}$.  We fix different values of $\alpha>3$ and choose $\beta>0$ such that $\Pr(p>p_\text{lo}|\alpha,\beta)=0.99$ under the Beta distribution for $p$.  We consider $\alpha=3.1,4,5,6,7$, and 8.  As before, we set $\pi(G_S)\propto \exp[-\gamma|E_S|]$ where $\gamma=\log(\hat{p}/(1-\hat{p}))$, with $\hat{p}=\alpha/(\alpha+\beta)$.

Table \ref{tab:results} shows posterior summaries for the estimated number $N$ of PWID in St.~Petersburg under each prior specification. The posterior mode, mean, standard deviation (SD), and 95\% posterior quantiles are shown.  \citet{Heimer2010Estimation} state that at least 30,000 cases of HIV in PWID have been reported; the 2.5\% quantile for $\alpha=3.1$ is just below this number.  Under this prior specification, increasing values of $\alpha$ decrease the prior mean of $p$, giving larger posterior estimates and variances of $N$.  The posterior mean $\E[N|\Y]$ is more sensitive than the mode to changes in $\alpha$ because it is strongly affected by the thickness of the right-hand tail of the posterior distribution.  We obtain posterior mode estimates between 53,000 and 210,000, which are generally compatible with that of \citet{Heimer2010Estimation}: posterior quantile intervals corresponding to $\alpha=3.1,4,5$, and $6$ contain their estimate $\hat{N}= 83,118$.  Setting $\alpha=8$ results in the highest estimates of over 200,000; estimates substantially larger than this may not be credible. The total number of people in St.~Petersburg is approximately 4.9 million, 
and \citet{Heimer2010Estimation} estimate the number who match the age range (20-45 years) characteristic of PWID as approximately 1.5 million.  The last two columns give the implied prevalence of injection drug use in both of these groups, computed using the posterior mean.  Posterior expectations and quantiles of $N$ in Table \ref{tab:results} are sensitive to the prior mean of $p$. The conditions required for the posterior distribution of $N$ to have finite variance necessitate informative priors for $p$ \citep{Kahn1987Cautionary}. Nevertheless, the estimates of the number $N$ of PWID in St.~Petersburg are in general agreement with those of \citet{Heimer2010Estimation} and span a range of reasonable values.

\begin{table}
\centering
\begin{tabular}{rrrrrrrrrr}
  \hline
  Prior  && \multicolumn{5}{c}{Population size $N$} && \multicolumn{2}{c}{Prevalence (\%)} \\ 
  \cline{1-1}\cline{3-7}\cline{9-10}
  $\alpha$ && Mode & Mean  & SD    & 2.5\% & 97.5\% && 20-45yrs & All \\ 
  \hline
  3.1 && 53797  & 89332  & 65919 & 28968 & 237986 && 5.9  & 1.8 \\ 
  4.0 && 77187  & 101692 & 48046 & 41394 & 216854 && 6.8  & 2.1 \\ 
  5.0 && 100307 & 141534 & 65067 & 58954 & 309433 && 9.4  & 2.9 \\ 
  6.0 && 125086 & 167100 & 69161 & 78403 & 345233 && 11.1 & 3.4 \\ 
  7.0 && 152588 & 209075 & 91636 & 95078 & 442319 && 13.9 & 4.3 \\ 
  8.0 && 170167 & 195711 & 64268 & 98111 & 351162 && 13.1 & 4.0 \\ 
   \hline
\end{tabular}
  \caption{Estimates of the number of people who inject drugs in St.~Petersburg, Russia from an RDS dataset of $n=813$ subjects. The prior for $p$ depends on $\tau$, defined in the text. Posterior means, standard deviations, and 2.5\% and 97.5\% quantiles are shown.  The last two columns show the approximate implied prevalence (\%) of injection drug use in 20-45 year-olds and for all residents of St.~Petersburg. }
\label{tab:results}
\end{table}

We also analyze the St.~Petersburg data using the SS-size method described by \citet{Handcock2014Estimating} and \citet{Handcock2015Estimating}.  Results are shown in Table 1 of the Supplementary Materials.  The SS-size model and the method proposed in this paper are quite different, but we have attempted to impose similar prior specifications so that the results are comparable between the two approaches. The posterior estimates from the SS-size method generally fall between 1000 and 4000 when the raw degrees $\bd$ are used, which is not within the feasible range for the number of PWID in St.~Petersburg.  Estimates increase to between 20,000 and 100,000 when subjects' reported degrees are ``imputed'' by the SS-size software.  Estimates under the SS-size model are highly sensitive to a user-specified maximum $N$ value.  Setting this maximum to $500,000$ results in the largest estimates.  The prior distribution imposed on $N$ does not seem to greatly affect the posterior estimates in the SS-size method. Estimates from the SS-size model using the raw degrees $\bd$ imply that the prevalence of injection drug use is between 0.09\% and 0.18\% for 20-45 year-olds and between 0.03\% and 0.06\% for all residents of St.~Petersburg, which is far lower than the known minimum prevalence based on the number of registered PWID, and the number of PWID known to be HIV-positive.  

\citet{Gile2011Improved}, \citet{Handcock2014Estimating}, \citet{Handcock2015Estimating}, and \citet{Gile2015Diagnostics} argue that degrees of subjects recruited by RDS should decrease as the sample accrues. 
One possible reason for the poor performance of the SS-size method in this dataset is that the time-ordered degrees do not adhere to this assumption.  The mean reported degree in the St.~Petersburg dataset is 10.26 with SD 8.5; the maximum reported degree is 200.  Figure 1 of the Supplementary Materials shows the reported degrees and a linear regression line overlaid.  To test whether the time-ordered sample of subjects' degrees decreases, we use the approach suggested by \citet{Gile2015Diagnostics} and regress the integers $1,\ldots,n$ on the observed degrees $\mathbf{d}$, ordered by the time of recruitment. We employ linear, Poisson, and $M$-estimation with Huber and bisquare weighting.  We fit these regression models using the full dataset of $n=813$ reported degrees and with the same dataset excluding one outlier subject who reported $d=200$. The results are shown in Table 2 of the Supplementary Materials.  The estimated slope coefficient is always small and positive.  There does not appear to be a significant negative trend in the reported degrees, and we conclude that average reported degrees do not decrease in this dataset.


\section{Discussion: models and assumptions}

\label{sec:dis}

We have presented a method for estimating the size of a hidden population from data collected during RDS surveys.  The proposed estimation method recovers the true value of $N$ accurately in simulations, and gives reasonable results in the application to estimate the number of PWID in St.~Petersburg.  The modeling approach relies on several assumptions about the social network connecting members of the target population and the RDS recruitment process.  In this section, we examine the basic assumptions underlying the method, and compare them to those made by \citet{Handcock2014Estimating} and \citet{Handcock2015Estimating} in deriving and justifying the SS-size estimator.

\subsection{The network}

In this paper, we have assumed that there exists an undirected social network $G=(V,E)$ connecting members of the hidden population, and Assumption \ref{assump:er} states that this network follows the \er distribution.  Human social networks are not usually well characterized by the \er model \citep{Watts1998Collective,Robins2001Network}.  However, the \er model has appealing properties in the context of hidden population size estimation: first, the likelihood \eqref{eq:erlik1} is simple and does not require calculation of a normalizing constant. Second, the \er model reflects our general ignorance about the social structure of hidden populations; setting $p=0.5$ gives the ``uniform'' distribution on graphs.  Third, because even small subgraphs can provide information about $N$ in the \er model, \eqref{eq:erlik1} does not require that the network be connected, nor that the sample take place in the giant component.  Finally, and most importantly, the \er model has proven to be empirically useful in a wide variety of population size estimation applications via the snowball sampling estimator \citep{Frank1994Estimating,David2002Estimating} and the network scale-up method \citep[e.g][]{Bernard2001Estimating}.  The success of these methods in real-world applications suggests that there may be some merit to the notion that certain kinds of acquaintanceships form somewhat independently and with common probability.  Moreover, the proposed method uses data from RDS surveys of hidden population members, whose within-group edge probabilities may be more homogeneous than between-group probabilities in the general population.

In contrast, the SS-size model of \citet{Handcock2014Estimating,Handcock2015Estimating} does not assume the existence of a network, and assigns degrees of unsampled vertices independently from a pre-specified parametric distribution.  This approach is unburdened by graph-theoretic constraints on the population network, since the set of population degrees drawn in this way need not correspond to the degree sequence of any graph \citep[see e.g.][]{Erdos1960Grafok}.  More importantly, inference under the SS-size model is not constrained by topological conditions imposed by the observed recruitment graph $G_R$ and the degrees in the subgraph of respondents, as in Definition \ref{def:compatibility}.  In the SS-size  method, network topology local to recruited vertices does not play a role in recruitment of the sample.  This lack of graphical constraints in the SS-size  model suggests a view of RDS recruitment that is not network-based: subjects' reported degrees might be regarded as surrogate measures of ``visibility'' in the population, and \citet{Handcock2014Estimating} and \citet{Handcock2015Estimating} takes sampling probability proportional to visibility.

\subsection{The recruitment process}

Assumption \ref{assump:samplingprob} states that the probability that a susceptible vertex is recruited depends only on its edges connecting to active recruiters, and does not depend on edges connecting to unsampled vertices.  In contrast, the SS-size method largely avoids modeling the recruitment process by assuming that sampling of subjects in RDS occurs with probability proportional to their total degree, without replacement.  This assumption has two important implications that highlight the difference between the SS-size model and the method developed in this paper.  First, the SS-size model of recruitment is not compatible with Assumption \ref{assump:samplingprob}, which states that the probability that a given vertex is recruited depends on the edges it shares with recruiters, and does not depend on edges that connect this vertex to other unrecruited vertices.  Indeed, under the SS-size model, network topology implied by the recruitment graph $G_R$ is irrelevant to the recruitment process and vertex degrees are treated as ``sizes'' in a ``probability proportional to size without replacement'' sampling scheme \citep[e.g.][]{Bickel1992Nonparametric}.  Second, the degrees of recruited subjects should \emph{decrease} over time as the sample accrues under the SS-size model.  We did not observe such a decrease in mean degree in the St.~Petersburg data (see the Supplementary Materials).  Nor did \citet[][Supplementary Materials]{Gile2015Diagnostics}, who find that in robust regression analyses of twelve separate RDS datasets, ``[s]urprisingly, we find little evidence of decreasing degree over time''.  

However, there is reason to believe that network topology matters in determining who can be recruited, that RDS sampling probability is not proportional to degree, and that degrees need not decrease during an RDS study. \citet{Crawford2014Graphical} argues that if RDS recruitments happen over edges of a population network, sampling probability has little to do with total degree.  Instead, the number of edges each potential subject (susceptible vertex) shares with recruiters determines their probability of being sampled in the next recruitment. Indeed, a potential subject who shares no edges with recruiters cannot be recruited, regardless of their degree.  Worse, sample sizes for RDS studies are usually set in advance, so a potential subject whose location in $G$ is more than $n$ edges from a seed can never be recruited, regardless of their degree.  When average degree does not decrease over the time-ordered sample, the assumptions underlying the SS-size  method may not be met, and the likelihood of the ordered degrees under the SS-size  process may not be informative for $N$.  

We also assume that per-susceptible-edge waiting time to recruitment is memoryless (Assumption \ref{assump:exp}), which provides a convenient marginalizing distribution over subgraphs $\widehat{G}_S$ in \eqref{eq:post1}.  To justify this assumption, we draw an analogy between the RDS recruitment process and the spread of an infectious disease on a population network.  The contact process between ``susceptible'' vertices and ``infective'' recruiters closely parallels models that have gained wide use in epidemiology.  The main difference is that recruiters can deplete their coupons in RDS, which renders them unable to recruit others.  The incentive for recruiting other participants in RDS may also provide some justification for exponential waiting times: the need for money may be essentially memoryless.


\section{Conclusion: RDS for population size estimation?}

RDS was not designed for population size estimation, and it should not be used if other options like census enumeration or capture-recapture are available and the assumptions necessary for their use are justified.  But RDS remains a popular survey method for good reason: it is a remarkably effective procedure for recruiting subjects who might otherwise be reluctant to participate in a research survey.  The lack of better methods for learning about hidden populations suggests to us that RDS will find continued use by epidemiologists and public health researchers in the future.  We have shown in this paper that by making some assumptions about the network and the nature of the RDS recruitment process, the observed data from an RDS study can provide useful information about the target population size.  The assumptions underlying this method may be justified when researchers believe that the population network exists, and subjects are recruited across its edges.




\section*{Supplementary Materials}

\section{Posterior distribution of $N$}

\label{app:post2}

Let $\pi(p) \propto p^{\alpha-1}(1-p)^{\beta-1}$ and $\pi(\lambda) \propto \lambda^{\eta-1} e^{-\xi\lambda}$ be prior distributions.  We find the posterior distribution of $N$ by marginalizing over subgraphs $\widehat{G}_S\in\mathcal{C}(G_R,\bd)$, $N$, and $p$,
\begin{equation}
  \Pr(N|\Y) \propto \pi(N)\sum_{\widehat{G}_S}\pi(\widehat{G}_S) \int_0^\infty  L(\widehat{G}_S,\lambda|\Y)\ \pi(\lambda) \int_0^1 L(N,p|\widehat{G}_S,\Y)\ \pi(p) \dx{p} \dx{\lambda} .
	\end{equation}
	The integral over $\lambda$ is
	\begin{equation}
		\begin{split}
    \int_0^\infty L(\widehat{G}_S,\lambda|\Y)\ \pi(\lambda) \dx{\lambda} &\propto \int_0^\infty \left(\prod_{j\notin M}\mathbf{s}_j\right) \lambda^{n-m} \exp[-\lambda \mathbf{s}'\w]\ \lambda^{\eta-1}e^{-\xi\lambda} \dx{\lambda} \\ 
			 &\propto \frac{ \prod_{j\notin M}\mathbf{s}_j }{(\mathbf{s}'\w+\xi)^{n-m+\eta}} \\
		\end{split}
	\end{equation}
and the integral over $p$ is 
	\begin{equation}
		\begin{split}
 \int_0^1 L(N,p|\widehat{G}_S,\Y)\ \pi(p) \dx{p} &\propto \left[\prod_{i=1}^n\binom{N-i}{d_i^u}\right] \int_0^1  p^{D^u}(1-p)^{N-\binom{n+1}{2}-D^u}\ p^{\alpha-1}(1-p)^{\beta-1} \dx{p} \\
 			 &= \left[\prod_{i=1}^n\binom{N-i}{d_i^u}\right] \int_0^1  p^{D^u+\alpha-1}(1-p)^{N-\binom{n+1}{2}-D^u+\beta-1} \dx{p} \\
         &\propto \left[ \prod_{i=1}^n \binom{N-i}{d_i^u} \right] \mathrm{B}\left(D^u+\alpha,nN-\binom{n+1}{2}-D^u+\beta\right)
\end{split}
\end{equation}
where the $d_i^u$'s are computed from $\widehat{G}_S$ and $\bd$, $D^u=\sum_{i=1}^n d_i^u$, and $\text{B}(\cdot,\cdot)$ is the Beta function.  The marginal posterior distribution of $N$ is therefore 
\begin{equation}
\Pr(N|\Y) \propto 
\pi(N)\sum_{\widehat{G}_S}
\frac{ \pi(\widehat{G}_S) \prod_{j\notin M}\mathbf{s}_j }{(\mathbf{s}'\w+\xi)^{n-m+\eta}} 
\left[ \prod_{i=1}^n \textstyle{\binom{N-i}{d_i^u}} \right] \textstyle{\mathrm{B}\left(D^u+\alpha,nN-\binom{n+1}{2}-D^u+\beta\right) }.
\label{eq:post}
\end{equation}


\section{Conditions for existence of moments of $\Pr(N|\Y)$}

\label{app:postproper}

The posterior mass function of $N$ is given by \eqref{eq:post}.  We seek sufficient conditions for the posterior mass function to be proper and to have finite first and second moments when $\pi(N)\propto N^{-c}$.  First, note that the sum over $\widehat{G}_S\in\mathcal{C}(G_R,\bd)$ is finite, so we ignore the sum over $\widehat{G}_S$ and consider the function 
\begin{equation}
  \Pr(N|\Y) \propto \left[ \prod_{i=1}^n \frac{(N-i)!}{(N-i-d_i^u)!} \right] \frac{\Gamma(nN-\binom{n+1}{2}-D^u+\beta)}{\Gamma(nN-\binom{n+1}{2}+\alpha+\beta)} N^{-c}
\end{equation}
where we have used the definition of the Beta function as a ratio of Gamma functions.  We first provide a bound for the product term, then the ratio of Gamma functions.  Each term in the product obeys the bound
\begin{equation}
\begin{split}
\frac{(N-i)!}{(N-i-d_i^u)!} &\leq \frac{(N-i)^{N-i+1/2} e^{-(N-i)+1}}{\sqrt{2\pi} (N-i-d_i^u)^{N-i-d_i^u+1/2} e^{-(N-i-d_i^u)} } \\
                            &\leq \frac{e^{-d_i^u+1}}{\sqrt{2\pi}} \left( \frac{N}{N-n-d_i^\text{max}} \right)^{N-i+1/2} (N-n-d_i^\text{max})^{d_i^u}  \\
\end{split}
\end{equation}
(via Stirling's approximation) where $d_i^\text{max} = \text{max}_i d_i$.  Then
\begin{equation}
  \begin{split}
    \prod_{i=1}^n \frac{(N-i)!}{(N-i-d_i^u)!} &\leq \text{const} \times \left( \frac{N}{N-n-d_i^\text{max}} \right)^{nN-\binom{n+1}{2}+n/2} (N-n-d_i^\text{max})^{D^u} .
\end{split}
\label{eq:prodbound}
\end{equation}
where the $d_i^u$'s are computed from $\widehat{G}_S$ and $\bd$ and $D^u=\sum_{i=1}^n d_i^u$.  Second, 
\begin{equation}
  \begin{split}
    \frac{\Gamma(nN-\binom{n+1}{2}-D^u+\beta)}{\Gamma(nN-\binom{n+1}{2}+\alpha+\beta)} &\leq  \frac{ (nN-\binom{n+1}{2} - D^u+\beta-1)^{nN-\binom{n+1}{2} - D^u+\beta-1/2} e^{-(nN-\binom{n+1}{2}-D^u+\beta-1)+1}}{\sqrt{2\pi} (nN-\binom{n+1}{2} +\alpha+\beta-1)^{nN-\binom{n+1}{2} +\alpha+\beta-1/2} e^{-(nN-\binom{n+1}{2}+\alpha+\beta-1)}} \\
    &= \left(\frac{nN-\binom{n+1}{2}-D^u+\beta-1}{nN-\binom{n+1}{2}+\alpha+\beta-1}\right)^{nN-\binom{n+1}{2}+\beta} \\
    &\qquad \times \frac{(nN-\binom{n+1}{2}-D^u+\beta-1)^{-D^u}}{(nN-\binom{n+1}{2}+\alpha+\beta-1)^{\alpha}} \frac{e^{D^u+\alpha+1}}{\sqrt{2\pi}}  \\
    &\leq \left(nN-\binom{n+1}{2}+\beta-1\right)^{-D^u-\alpha} \frac{e^{D^u+\alpha+1}}{\sqrt{2\pi}} .
  \end{split}
  \label{eq:Gammabound}
\end{equation}
Combining \eqref{eq:prodbound} and \eqref{eq:Gammabound}, we have 
\begin{equation}
  \begin{split}
    \Pr(N|\Y) &\leq \text{const} \times \left( \frac{N}{N-n-d_i^\text{max}} \right)^{nN-\binom{n+1}{2}+n/2} (N-n-d_i^\text{max})^{D^u}  \\
    &\qquad \times \left(nN-\binom{n+1}{2}+\beta-1\right)^{-D^u-\alpha} N^{-c} \\
    &=  \text{const} \times \left( \frac{N}{N-n-d_i^\text{max}} \right)^{nN-\binom{n+1}{2}+n/2} \left(\frac{N-n-d_i^\text{max}}{nN-\binom{n+1}{2}+\beta-1}\right)^{D^u}\\
    &\qquad \times \left(nN-\binom{n+1}{2}+\beta-1\right)^{-\alpha} N^{-c} 
  \end{split}
\end{equation}
The first term converges to one, the second to a constant that does not depend on $N$, while the last two terms dominate in the right-hand tail, and for large $N$ we have
\begin{equation}
  \begin{split}
   \Pr(N|\Y) &\approx  \left(nN-\binom{n+1}{2}+\beta-1\right)^{-\alpha} N^{-c} \\
             &\propto N^{-(\alpha+c)} .
  \end{split}
\end{equation}
It follows that a sufficient condition for the posterior distribution to be proper is $\alpha+c>1$.  The condition $\alpha+c>2$ ensures that the posterior mean exists, and $\alpha+c>3$ ensures that the second moment exists, and hence the posterior variance.


\section{Gibbs sampling for $G_S$ and $N$}

\label{app:gibbs}

\subsection{Sampling $G_S$ given $N$}

\citet{Crawford2014Graphical} describes a procedure for drawing a proposal subgraph $\widehat{G}_S$ uniformly from the set of compatible subgraphs $\mathcal{C}(G_R,\bd)$.  Let $m=|M|$ be the number of seeds.  The conditional posterior distribution of $G_S$ is 
\begin{equation}
	\Pr(\widehat{G}_S|N,\Y) \propto \frac{\prod_{j\notin M} \mathbf{s}_j}{(\mathbf{s}'\w+\xi)^{n-m+\eta}} \left[ \prod_{i=1}^n \textstyle{\binom{N-i}{d_i^u}} \right] \textstyle{\mathrm{B}\left(D^u+\alpha,nN-\binom{n+1}{2}-D^u+\beta\right)} \pi(\widehat{G}_S)
\end{equation}
where $\mathbf{s}$, $\bd_i^u$, and $D^u$ are computed using $\widehat{G}_S\in\mathcal{C}(G_R,\bd)$.  

Suppose $G_S=(V_S,E_S)$ is the current estimate of the recruitment-induced subgraph.  We propose a new subgraph by adding or removing an edge from this graph. To draw a new sample from $\mathcal{C}(G_R,\bd)$, we select vertices $i$ and $j$, with $i\neq j$ at random.  Then if $\{i,j\}\notin E_S$, $\bu_i>0$, and $\bu_j>0$, we propose to add the edge $\{i,j\}$ to $E_S$.  If $\{i,j\}\in E_S$ and $\{i,j\}\notin E_R$, we propose to remove the edge $\{i,j\}$ from $E_S$.  Otherwise, we select a different $\{i,j\}$ and try again.  The vector of the number of susceptible vertices just before each recruitment is $\mathbf{s}=\text{lowerTri}(\A\C)'\mathbf{1} + \C'\bu$ using the current subgraph estimate $G_S$ and let $\mathbf{s}^+$ and $\mathbf{s}^-$ be the corresponding vectors obtained by adding or removing an edge between $i$ and $j$.  It is not necessary to compute $\mathbf{s}$ via matrix multiplication.  Instead, \citet{Crawford2014Graphical} provides the update expressions
\begin{equation}
  \begin{split}
    \mathbf{s}_k^+ &= \mathbf{s}_k - \indicator{k>j}C_{ik} - C_{jk} \\
    \mathbf{s}_k^- &= \mathbf{s}_k + \indicator{k>j}C_{ik} + C_{jk} ,
  \end{split}
  \label{eq:skrem}
\end{equation}
for $k=1,\ldots,n$.  Now let $t_i^*$ be the time at which vertex $i$ used all its coupons or the end of the study, whichever came first.  Then the change in total edge-time is given by
\begin{equation}
  \begin{split}
    \mathbf{s}^{+'}\w &= \mathbf{s}'\w - (t_i^* - \text{min}(t_j,t_i^*) + t_j^* - t_j)  \\
    \mathbf{s}^{-'}\w &= \mathbf{s}'\w + (t_i^* - \text{min}(t_j,t_i^*) + t_j^* - t_j) .
  \end{split}
\end{equation}
Using these expressions, the ratio of posterior probabilities for $N$ reduces to a simple form.  To illustrate, suppose we wish to add the edge ${i,j}$ to $G_S=(V_S,E_S)$, where $\{i,j\}\notin E_S$, $\bu_i\geq 1$, and $\bu_j\geq 1$.  For a proposal $G_S^+=(V_S,E_S^+)$ identical to $G_S$ except that $\{i,j\}\in E_S^+$, $\bu_i^+ = \bu_i-1$, and $\bu_j^+ = \bu_j-1$, the ratio is
\begin{equation} 
  \frac{\Pr(G_S^+|N,\Y)}{\Pr(G_S|N,\Y)} = \left( \prod_{j\notin M} \frac{\mathbf{s}_j^+}{\mathbf{s}_j} \right) \left( \frac{\mathbf{s}'\w+\xi}{\mathbf{s}^{+'}\w+\xi}\right)^{n-m+\eta} \frac{d_j^u}{N-j-d_j^u+1} \cdot  \frac{nN-\binom{n+1}{2}-D^u+\beta}{D^u-1+\alpha}\cdot \frac{\pi(G_S^+)}{\pi(G_S)}  .
\end{equation}
To illustrate the ratio for removing the edge ${i,j}$, suppose $G_S=(V_S,E_S)$ has $\{i,j\}\in E_S$ and $\{i,j\}\notin E_R$.  For a proposal $G_S^-=(V_S,E_S^-)$ identical to $G_S$ except that $\{i,j\}\notin E_S^-$, $\bu_i^- = \bu_i+1$, and $\bu_j^- = \bu_j+1$, the ratio is
\begin{equation} 
  \frac{\Pr(G_S^-|N,\Y)}{\Pr(G_S|N,\Y)} = \left( \prod_{j\notin M} \frac{\mathbf{s}_j^+}{\mathbf{s}_j} \right) \left( \frac{\mathbf{s}'\w+\xi}{\mathbf{s}^{-'}\w+\xi}\right)^{n-m+\eta} \frac{N-j-d_j^u}{d_j^u+1} \cdot  \frac{D^u+\alpha}{nN-\binom{n+1}{2}-D^u-1+\beta} \cdot \frac{\pi(G_S^-)}{\pi(G_S)} .
\end{equation}
Suppose $G_S^*$ is the proposal graph and let $\Pr(G_S^*|G_S)$ be the probability of proposing $G_S^*$ from $G_S$, with $N$ fixed.  To decide whether to accept $G_S^*$, we form the Metropolis-Hastings acceptance probability, 
\begin{equation}
  \rho = \min\left\{1,\ \frac{\Pr(G_S^*|N,\Y)}{\Pr(G_S|N,\Y)} \frac{\Pr(G_S|G_S^*)}{\Pr(G_S^*|G_S)} \right\} .
  \label{eq:mhG}
\end{equation}
The form of $\Pr(G_S^*|G_S)$ is given by \citet{Crawford2014Graphical}.


\subsection{Sampling $N$ given $G_S$}

\label{app:Nhat}

The posterior distribution of $N$ conditional on a given compatible subgraph $G_S$ is 
\begin{equation}
  \Pr(N|G_S,\Y) \propto \left[ \prod_{i=1}^n \binom{N-i}{d_i^u} \right] \mathrm{B}\left(D^u+\alpha,nN-\binom{n+1}{2}-D^u+\beta\right) \pi(N) 
  \label{eq:postN}
\end{equation}
Although this conditional distribution does not have a standard form, we can derive a close approximation using the negative binomial distribution when $\Pr(N|G_S,\Y)$ has a mode.  Let $d_1^u,\ldots,d_n^u$ be the number of pendant edges emanating from each sampled vertex at the moment they are recruited, calculated from $G_S$. Suppose for now that $N$ is continuous-valued.  We can calculate analytic derivatives of $\ell(N)=\log \Pr(N|G_S,\Y)$ as follows:
\begin{equation}
  \begin{split}
  \pd{\ell}{N} &= \left[ \sum_{i=1}^n \psi(N-i+1) - \psi(N-i-d_i^u+1)\right] \\ 
  &\qquad + \left[\psi\left(nN-\binom{n+1}{2} - D^u + \beta\right) - \psi\left(nN-\binom{n+1}{2} +\alpha + \beta\right)\right] n - \frac{c}{N} \\
  \pdm{\ell}{N}{2} &= \left[ \sum_{i=1}^n \psi^{(1)}(N-i+1) - \psi^{(1)}(N-i-d_i^u+1)\right] \\ 
  &\qquad + \left[\psi^{(1)}\left(nN-\binom{n+1}{2} - D^u + \beta\right) - \psi^{(1)}\left(nN-\binom{n+1}{2} + \alpha + \beta\right)\right] n^2 + \frac{c}{N^2} \\
\end{split}
\end{equation}
where $\psi(x)=\pd{\log\Gamma(x)}{x}$ is the digamma function and $\psi^{(1)}(x) = \pdm{\log\Gamma(x)}{x}{2}$ is the polygamma function. Let $\hat{N}=\text{argmax}_N \ell(N)$ be the mode of $\Pr(N|G_S,\Y)$ and let 
\begin{equation}
  v=\left( -\left.\pdm{\ell}{N}{2}\right|_{N=\hat{N}} \right)^{-1}
 \end{equation} 
 be an approximation to the variance.  To draw from $\Pr(N|G_S,\Y)$ we employ a proposal distribution to generate a candidate $N^*$ and use a Metropolis-Hastings correction to draw from the relevant conditional posterior.  We will use $\hat{N}$ and $v$ to construct a proposal distribution for $N$ given $G_S$.  Consider $N^* \sim \text{NegBin}(\hat{N},r)$, where we have parameterized the negative binomial distribution by its mean and size $r$.  The variance of the proposal distribution under this parameterization is $N + N^2/r$, so to achieve a proposal variance of $v$, where $v>N$, set $r=N^2/(v-N)$.  The proposal distribution is
\begin{equation}
  \Pr(N^*=k|\hat{N}) = \left. \left(\frac{\hat{N}}{r+\hat{N}}\right)^k \frac{\Gamma(r+k)}{k!} \middle/ \sum_{j=N_\text{min}}^\infty \left(\frac{\hat{N}}{r+\hat{N}}\right)^j \frac{\Gamma(r+j)}{j! \Gamma(r)} \right. , 
  \label{eq:Nstar}
\end{equation}
where we have normalized by the probability that $N^*\geq N_\text{min}$.  Then the Metropolis-Hastings ratio for the proposal $N^*$ conditional on $G_S$ is 
\begin{equation}
  \rho = \min\left\{1,\ \frac{\Pr(N^*|G_S,\Y)}{\Pr(N|G_S,\Y)} \frac{\Pr(N|\hat{N})}{\Pr(N^*|\hat{N})} \right\} .
  \label{eq:mhN}
\end{equation}
The infinite sum in the denominator of \eqref{eq:Nstar} cancels in the ratio \eqref{eq:mhN}. 


\section{An approximation for prior elicitation}

\label{app:elicitation}

Suppose we wish to find values of $\alpha$ and $\beta$ that place the prior mean of $N$ approximately equal to $\hat{N}$, a prior estimate of $N$.  Recall that $d_i^u$ follows the Beta-Binomial distribution, and let $\bar{p}=\alpha/(\alpha+\beta)$.  Then $\E[d_i^u] = (N-i)\bar{p}$ and 
\begin{equation}
  \E\left[\sum_{i=1}^n d_i^u\right] = \bar{p} \left( nN - \binom{n+1}{2} \right) .
\end{equation}
Equating observed and expected values of $d_i^u$ and rearranging, we have an estimator for $N$ given $\bar{p}$, 
\begin{equation}
  \tilde{N} = \frac{n+1}{2} + \frac{1}{\bar{p} n} \sum_{i=1}^n d_i^u 
  \label{eq:Nest}
\end{equation}
or an estimator for $\bar{p}$ given $N$, 
\begin{equation}
  \tilde{p} = \frac{\sum_{i=1}^n d_i^u}{nN-\binom{n+1}{2}}.
  \label{eq:pest}
\end{equation}
Now let $N=\hat{N}$ in \eqref{eq:pest}.  Since $G_S$ is not directly observed in an RDS study, the $d_i^u$'s are not available.  However, we can place a sharp lower bound on the numerator of \eqref{eq:pest} by conditioning on the observed degrees.  Let $r_i$ be the number of subjects recruited by subject $i$ over the course of the study.  The number of edges belonging to vertex $i$ connecting to unrecruited vertices at the time of its recruitment cannot be smaller than $r_i$.  But at most $i-1$ edges of $i$ can connect to already-recruited vertices, so $\text{max}\{r_i, d_i-(i-1)\}$ is a lower bound for $d_i^u$.  Recall that $M$ is the set of seeds.  Then we have the lower bound
\begin{equation}
  \text{max}\{r_i, d_i-i+1\} \leq d_i^u 
\end{equation}
This leads us to a lower bound for $\bar{p}$ that depends only on $\hat{N}$ and information contained in $\mathbf{d}$ and $G_R$:  
\begin{equation}
  \frac{\sum_{i=1}^n \text{max}\{r_i, d_i-i+1\} }{n\hat{N}-\binom{n+1}{2}} \leq \tilde{p}
  \label{eq:pbounds}
\end{equation}
Let $p_\text{lo}$ denote this lower bound.  One strategy for prior elicitation is to restrict the prior distribution of $p$ so that $\Pr(p<p_\text{lo})$ is small.  We therefore fix $\alpha$ and find $\beta$ so that $\Pr(p>p_\text{lo}|\alpha,\beta)=0.99$.


\section{Results of SS-size method on the St.~Petersburg dataset}

\label{app:ss}

\begin{figure}
  \centering
  \includegraphics[width=\textwidth]{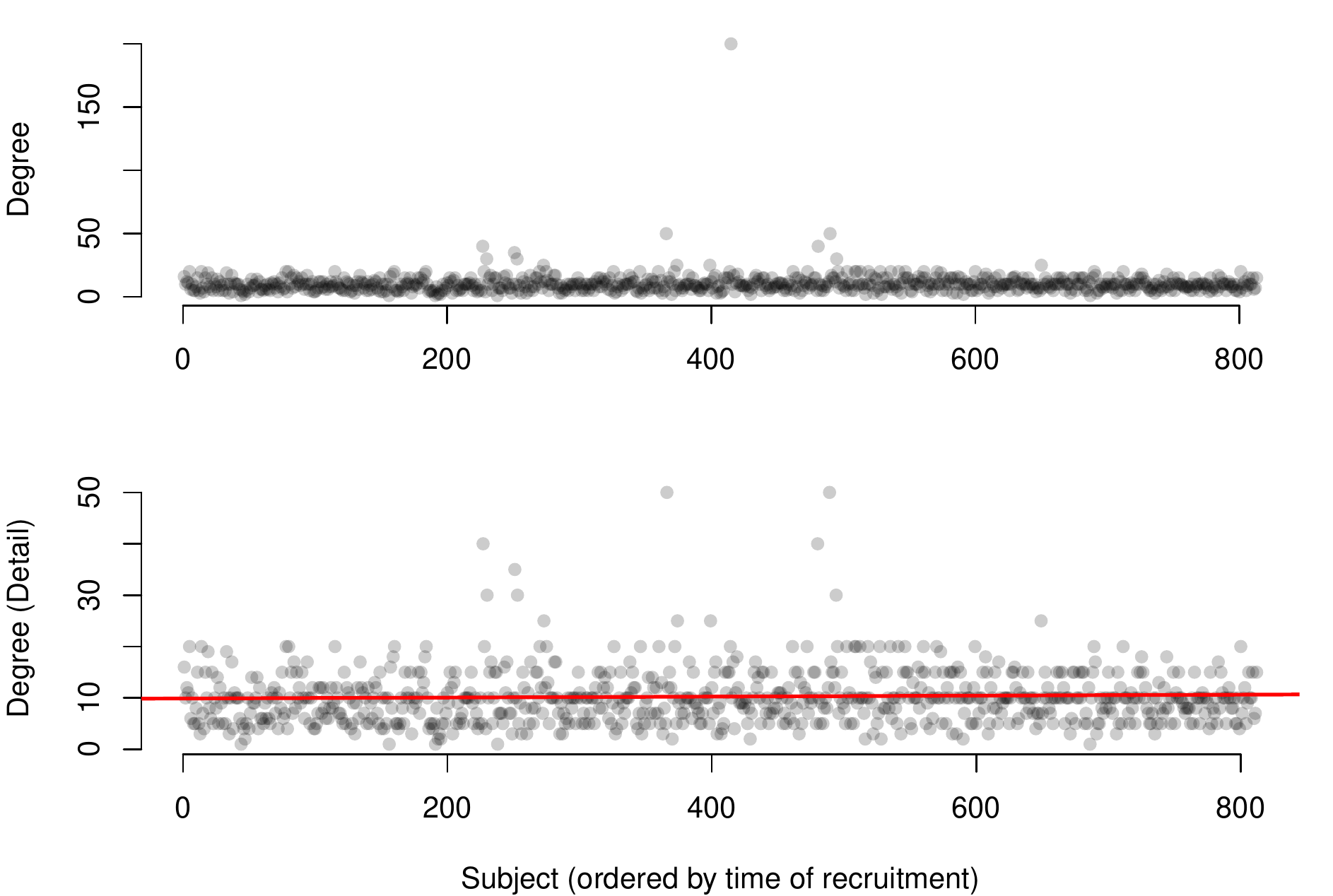}
  \caption{Degrees of recruited subjects in the St.~Petersburg study of PWID. The mean reported degree is 10.26, with SD 8.5.  One subject reported degree 200. The linear regression line, with slightly positive slope, is overlaid. }
  \label{fig:degrees}
\end{figure}

Table \ref{tab:rdsanalyst} shows the estimated number of PWID in St.~Petersburg using the SS-size method implemented in the ``size'' package \citep{Handcock2014Estimating,Handcock2015Estimating}.  Table \ref{tab:degrees} shows the results of regression analyses to determine whether the reported degrees in the St.~Petersburg data decrease over time as the sample accrues.  Figure \ref{fig:degrees} shows the reported degrees.  

\begin{table}
\centering
\begin{tabular}{llllllllllllll}
  \hline
  \multicolumn{3}{c}{Prior Parameters} && \multicolumn{3}{c}{Estimates} && \multicolumn{2}{c}{Implied Prevalence}\\
  \cline{1-3} \cline{5-7} \cline{9-10}
  $n/N$ & Max $N$ & Size &&  Mean & 2.5\% & 97.5\% && 20-45yrs & All  \\ 
  \hline
  Beta($\gamma=1$) & 200000 & raw && 2744 & 2209 & 3206 && 0.18\% & 0.06\%  \\ 
  Beta($\gamma=5$) & 200000 & raw && 2750 & 2209 & 3206 && 0.18\% & 0.06\%  \\ 
  Beta($\gamma=10$)& 200000 & raw && 2733 & 2209 & 3206 && 0.18\% & 0.06\%  \\ 
  \hline
  Beta($\gamma=1$) & 500000 & raw && 2072 & 1812 & 2312 && 0.14\% & 0.04\%  \\ 
  Beta($\gamma=5$) & 500000 & raw && 2064 & 1812 & 2312 && 0.14\% & 0.04\%  \\ 
  Beta($\gamma=10$)& 500000 & raw && 2058 & 1812 & 2312 && 0.14\% & 0.04\%  \\ 
  \hline
  Beta($\gamma=1$) & 200000 & imputed && 41948 & 12178 & 98911 && 2.80\% & 0.86\%  \\ 
  Beta($\gamma=5$) & 200000 & imputed && 43392 & 13574 & 97515 && 2.89\% & 0.89\%  \\ 
  Beta($\gamma=10$)& 200000 & imputed && 46464 & 12178 & 99509 && 3.10\% & 0.95\%  \\ 
  \hline
  Beta($\gamma=1$) & 500000 & imputed && 28005 & 7309 & 62274 && 1.87\% & 0.59\%  \\ 
  Beta($\gamma=5$) & 500000 & imputed && 22315 & 6310 & 50782 && 1.49\% & 0.46\%  \\ 
  Beta($\gamma=10$)& 500000 & imputed && 27096 & 7309 & 61775 && 1.81\% & 0.55\%  \\ 
  \hline
  Flat($\gamma=1$) & 200000 & raw && 1436 & 1212 & 1611 && 0.10\% & 0.03\%  \\ 
  Flat($\gamma=5$) & 200000 & raw && 1433 & 1212 & 1611 && 0.10\% & 0.03\%  \\ 
  Flat($\gamma=10$)& 200000 & raw && 1432 & 1212 & 1611 && 0.10\% & 0.03\%  \\ 
  \hline
  Flat($\gamma=1$) & 500000 & raw && 1350 & 1313 & 1812 && 0.09\% & 0.03\%  \\ 
  Flat($\gamma=5$) & 500000 & raw && 1354 & 1313 & 1812 && 0.09\% & 0.03\%  \\ 
  Flat($\gamma=10$)& 500000 & raw && 1351 & 1313 & 1812 && 0.09\% & 0.03\%  \\ 
  \hline
  Flat($\gamma=1$) & 200000 & imputed && 27351 & 2807 & 87945  && 1.82\% & 0.56\%   \\ 
  Flat($\gamma=5$) & 200000 & imputed && 31822 & 2807 & 105690 && 2.12\% & 0.65\%  \\ 
  Flat($\gamma=10$)& 200000 & imputed && 40331 & 2807 & 126626 && 2.69\% & 0.82\%  \\ 
  \hline
  Flat($\gamma=1$) & 500000 & imputed && 26440 & 2812 & 88258  && 1.76\% & 0.54\%  \\ 
  Flat($\gamma=5$) & 500000 & imputed && 38355 & 3311 & 128733 && 2.56\% & 0.78\%   \\ 
  Flat($\gamma=10$)& 500000 & imputed && 90623 & 4311 & 295628 && 6.04\% & 1.85\%   \\ 
   \hline
\end{tabular}
  \caption{Estimates from the ``size'' software of the number of people who inject drugs in St.~Petersburg, Russia. 
  We obtained posterior estimates under the flat (uniform) prior and Beta prior for the sample proportion $n/N$.  The Conway-Maxwell-Poisson (CMP) distribution is the prior for the population degree distribution $f(d|\eta)$.  We obtain results under two values for the maximum possible $N$: 200,000 and 500,000.  We set the prior mean of $N$ to 83118 and the prior standard deviation to $\gamma\times 5799$ where $\gamma\ge 1$, based on the estimate by \citep{Heimer2010Estimation}.  By increasing $\gamma$ to 5, 10, and 20, we obtain priors for $N$ with greater variance.  We set the mean, standard deviation, and maximum of the degree distribution equal to their sample counterparts. }
  \label{tab:rdsanalyst}
\end{table}

\begin{table}
\centering
\begin{tabular}{lccccccc}
  & \multicolumn{3}{c}{All degrees} && \multicolumn{3}{c}{Excluding $d=200$} \\ \cline{2-4} \cline{6-8}
  Method & Slope       & SE        & $p$-value && Slope & SE & $p$-value \\
  \hline 
        Linear & $9.2\times 10^{-4}$ & $1.3\times 10^{-3}$ & 0.47 && $8.9\times 10^{-4}$ & $7.9\times 10^{-4}$ & 0.26 \\
       Poisson & $9.0\times 10^{-5}$ & $4.7\times 10^{-5}$ & 0.05 && $8.9\times 10^{-5}$ & $4.7\times 10^{-5}$ & 0.06 \\
$M$ (Huber)    & $1.2\times 10^{-3}$ & $6.7\times 10^{-4}$ &      && $1.2\times 10^{-3}$ & $6.7\times 10^{-4}$ &   \\
$M$ (Bisquare) & $2.3\times 10^{-3}$ & $6.7\times 10^{-4}$ &      && $1.3\times 10^{-3}$ & $6.7\times 10^{-4}$ &   \\
  \hline  \\
\end{tabular}
\caption{Regression results for the slope of the time-ordered sample of degrees in the St.~Petersburg data.  The SS method of \citet{Handcock2014Estimating} and \citet{Handcock2015Estimating} assumes that the average degree of recruited subjects decreases as the sample accrues.  We fit linear, Poisson, and $M$ estimates with Huber and bisquare weighting for the full set of degrees, and with one outlier $(d=200)$ removed.  Estimated slope for the regression line is always positive, indicating that degrees appear to increase in this sample.}
\label{tab:degrees}
\end{table}

\spacingset{1}
\bibliographystyle{agsm}
\bibliography{rds}

\end{document}